\documentclass[twocolumn,pra,aps,showpacs]{revtex4}
\usepackage{epsfig}

\begin{document}
\title{Competition between Factors Determining Bright versus Dark Atomic States within a Laser Mode}

\author{Bryan~Hemingway}
\author{T.~G.~Akin}
\author{Steven~Peil}
\affiliation{Clock Development Division, Precise Time Department, United States Naval Observatory, Washington, DC 20392}
\author{J.~V.~Porto}
\affiliation{Joint Quantum Institute, National Institute of Standards and Technology and the University of Maryland, College Park, Maryland 20742 USA}

\date{\today}

\begin{abstract}

We observe bimodal fluorescence patterns from atoms in a fast atomic beam when the laser excitation occurs in the presence of a magnetic field and the atoms sample only a portion of the laser profile.  The behavior is well explained by competition between the local intensity of the laser, which tends to generate a coherent-population-trapping (CPT) dark state in the $J=1$ to $J'=0$ system, and the strength of an applied magnetic field that can frustrate the CPT process. This work is relevant for understanding and optimizing the detection process for clocks or other coherent systems utilizing these transitions and could be applicable to {\em in situ} calibration of the laser-atom interaction, such as the strength of the magnetic field or laser intensity at a specific location.

Keywords: fluorescence; dark state; CPT

\end{abstract}

\maketitle


\section{Introduction}

Electric-dipole coupled atomic transitions are widely used for quantum-state measurement~\cite{measurement} and momentum transfer~\cite{phillips} in laser interactions because of the high rate of photon scattering when driven with a resonant optical field.  Sustained driving of these strong transitions can be straightforward when the angular momentum of the higher-energy state is greater than that of the lower state, $J' > J$, supporting a cycling transition between two sublevels.  When this condition is not met, an atom can be optically pumped to a state that no longer couples to the polarization of the excitation laser, or, if multiple polarizations are present to couple all of the states in the lower manifold, destructive interference among the various transition pathways can inhibit excitation via coherent population trapping (CPT)~\cite{arimondo}.  In either case a dark state arises and photon scattering ceases.

Each of these types of dark state has been used in a variety of applications, from eliminating heating due to spontaneous emission in sub-recoil cooling~\cite{subrecoil} to creating microwave resonances for atomic clocks~\cite{CPTclock}. But the inhibition of desirable scattering is a negative consequence of dark states for state detection~\cite{ion} and laser manipulation of external degrees of freedom.  In the case of CPT, the dark state can be frustrated by application of a magnetic field, resulting in excited-state population and photon scattering~\cite{berkeland}. The field introduces time evolution of the magnetic sublevels, disrupting the steady-state superposition that suppresses laser excitation.

The success or failure in suppressing a CPT dark state with a magnetic field is determined by the relative strength of the Zeeman frequency shift of the sublevels and the Rabi frequency, $\Omega$, characterizing the strength of the coherent interaction between the lower- and higher-energy states~\cite{rms}. For a transition used for detection in an atomic clock or other coherent system, the signal-to-noise ratio (SNR) is maximized by collecting the most photons possible, necessitating a large enough magnetic field to optimize the excited state population and photon scattering rate.  This needs to be balanced with the size of the magnetic-field induced frequency shifts impacting the accuracy, stability or more generally the coherence of the system.  This is a known problem in trapped-ion optical clocks~\cite{ion,ion2}.  More recently optical clocks based on beams of neutral alkaline-earth atoms are utilizing (atom-)background-free detection on a transition with $J'<J$ and therefore vulnerable to CPT~\cite{ludlow, Taylor}.

In the case of atoms driven in a uniform magnetic field by a Gaussian laser mode, the relative strengths of the Zeeman shift and $\Omega$ can vary over the changing local intensity of the spatial mode profile, and both photon scattering and dark-state behavior may be present.  Coherent processes are often obscured, however, in an ensemble of atoms driven by an optical field due to the varying Rabi frequency across both dimensions of the spatial mode.  Mitigating this by simply enlarging the beam comes at the expense of intensity, so past efforts at observing coherent processes have used tailored flat-top laser modes to produce a uniform intensity~\cite{Lim, Reetz}.  Efforts to remove the effects of averaging over different intensities could also benefit from imaging the result of the interaction to recover some spatial information rather than collecting scattered light from atoms sampling different $\Omega$s on a single-pixel photodetector.

Here we present observation of the competition between factors leading to strong photon scattering and CPT along the profile of a laser mode.  The atomic transition subject to this behavior is only accessible after initially transferring population from the (global) ground state to a metastable state with a separate laser excitation; the relative size of the spatial modes used for the two transitions allows us to sample just a fraction of the laser mode of interest and remove much of the averaging that would occur if sampling the entire profile.

\section{Experiment}

\begin{figure}
\includegraphics[width=0.5\textwidth]{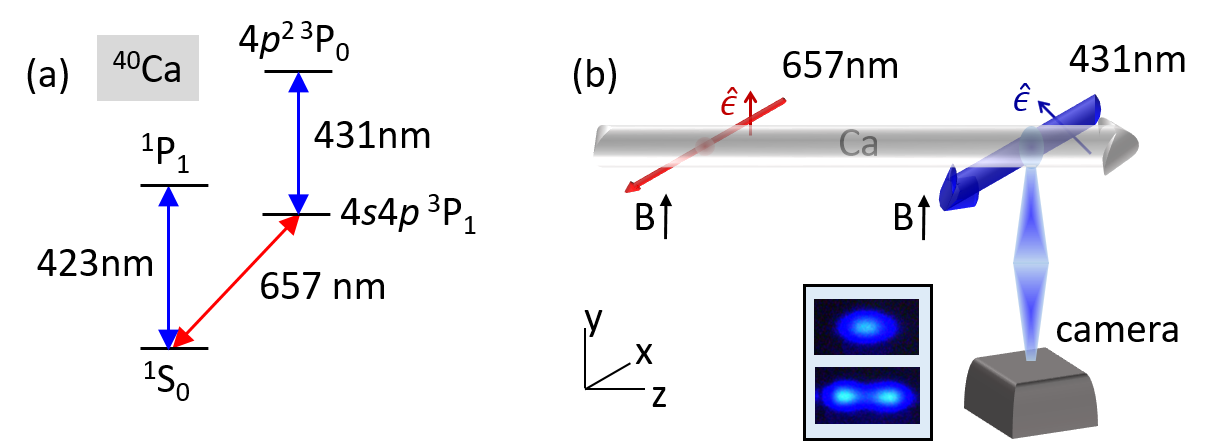}
\caption{(Color online.) (a)~Calcium energy level diagram, showing the narrow 657~nm transition and strong 423~nm and 431~nm transitions. (b)~Illustration of experimental arrangement. Atoms interact with linearly polarized 657~nm laser light in the presence of a magnetic field.  The size of the 657~nm mode is smaller than the spatial extent of the atomic beam, and smaller than the mode of the 431~nm laser in the subsequent interaction region.  The 431~nm light is linearly polarized at 45~degrees with respect to vertical ($y$), providing coupling among all Zeeman sublevels. A CCD camera below captures fluorescence images in the $xz$ plane. Inset: Example fluorescence images obtained for laser power of $\sim 10$~mW (top) and $\sim 500$~mW (bottom).} \label{f.illustration}
\end{figure}

Our system consists of a thermal (no laser cooling) beam of neutral calcium atoms interacting with lasers of two different wavelengths at two different positions, as illustrated in Fig.~\ref{f.illustration}.  Like other alkaline-earth atoms, calcium has an inter-combination transition resulting from a change in spin state of the two valence electrons. This weak $^1{\rm S_0}-4s4p~^3{\rm P_1}$ transition corresponds to a wavelength of 657~nm and has a natural width of about 400~Hz.  It is useful as an optical frequency reference and is particularly well suited for spectroscopy with a fast atomic beam.  Additionally, there is a strong transition with a wavelength of 431~nm that couples the $4s4p~^3{\rm P_1}$ metastable state, populated by atoms excited by 657~nm light, to a $4p^{2}~^3{\rm P_0}$ state~\cite{ludlow, Taylor}.

The atomic beam is generated by heating calcium granules in a molecular-beam epitaxy (MBE) cell to 650$^{\circ}$C, creating an effusive source with most probable velocity $v\sim500$~m/s.  The beam traverses a zone where the $^1{\rm S_0}-4s4p~{}^3{\rm P_1}$ inter-combination transition is driven with $\sim 10$~mW of 657~nm laser light. This light is linearly polarized vertically (along $y$ in Fig.~\ref{f.illustration}(b)), which selectively excites the $m_J=0$ to $m_J\prime=0$ transition in the presence of a $\sim500~\mu$T (5~G) vertical magnetic field. The narrow 400~Hz resonance is not resolved for fast atoms, where the time spent in the path of the laser limits the achievable linewidth to about 500~kHz for a mm-wide beam.

The lifetime of the $4s4p~^3{\rm P_1}$ state is on order of 400~$\mu$s, and over the course of a 1~m long beamline only a fraction of the atoms excited will decay back to the $^1{\rm S_0}$ ground state, yielding a detection signal of fewer than one photon per excited atom.  Instead, those atoms excited to the metastable $4s4p~^3{\rm P_1}$ state can be driven to the short-lived, 5~ns lifetime, $4p^2~^3{\rm P_0}$ state using 431~nm light.  This strong transition enables many blue photons to be scattered by a single atom originally excited by 657~nm light.  This transition is driven in the second interaction region shown in Fig.~\ref{f.illustration}(b). The 431~nm light is a collimated beam with an approximately elliptical spatial mode with horizontal ($z$) and vertical ($y$) sizes (1/e$^2$ diameter) of 2.5~mm and 4~mm.  As in the 657~nm excitation region, a $\sim500~\mu$T (5~G) vertical magnetic field provides a quantization axis.  Fluorescence in this region is maximized for a linear polarization at $\sim45$ degrees from vertical.  Vertical linear polarization is required to drive atoms from the $4s4p~^3{\rm P_1}$ $m_J=0$ state to the $4s^2~^3{\rm P_0}$ level.  Circularly polarized light can then drive $\sigma+$ and $\sigma-$ transitions connecting $4s4p~^3{\rm P_1}$ $m_J=-1$ and $m_J=1$ to $4s^2~^3{\rm P_0}$.

Laser-induced fluorescence from the 431~nm transition can be collected on a detector, such as a photomultiplier tube (PMT) or photodiode. In order to acquire spatial information, the fluorescence can also be imaged, as seen in the inset to Fig.~\ref{f.illustration}(b), where we show two different frames taken with a conventional CCD color video camera.  As illustrated, we will refer to a coordinate system in which the $k$-vectors of the laser beams point along $x$, the atomic velocity is along $z$, and we image the fluorescence from the atoms along $y$. The fluorescence images then capture the spatial pattern in the $xz$ plane, with the length of the pattern along $z$ determined by the spatial mode (and intensity) of the 431~nm laser and the thickness along $x$ determined by the velocities in the atomic beam with transverse Doppler shifts of $\pm250$~kHz or less, which are resonant with the 657~nm light~\cite{width}.

\begin{figure}
\includegraphics[width=0.5\textwidth]{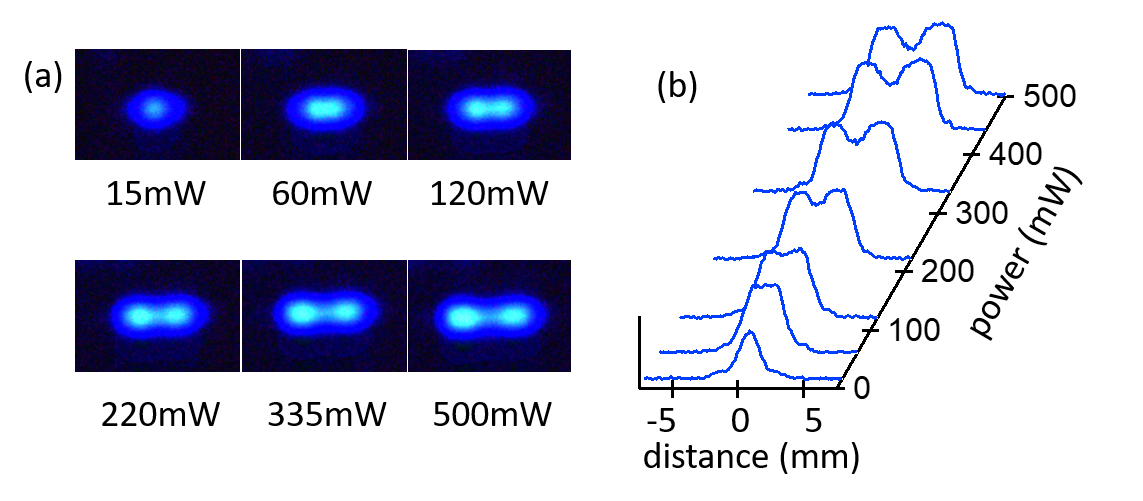}
\caption{(Color online.) (a) True-color fluorescence images acquired at different 431~nm laser powers.  The length of the image ($z$ direction) is determined by the dimension of the 431~nm laser profile along the atomic beam direction (and by the laser power).  Because these pictures are captured from below ($y$ direction), the height of the image ($x$ direction) is related to the frequency resolution of the 657~nm excitation, which is limited to $\sim$500~kHz. (b) Waterfall plot of 1D cross-sections of the fluorescence images.  One curve for a power of 440~mW is included here but not shown as an image in (a).  Some saturation exists in the blue channel of the RGB images acquired at the higher powers.} 
\label{f.images}
\end{figure}

Illumination of an atomic sample with a Gaussian beam typically results in a similar pattern of fluorescence, as long as the intensity is below saturation.  This is the case for the top image in the inset of Fig.~\ref{f.illustration}(b).  We observe that when the power in the 431~nm beam is increased beyond a certain value, the imaged fluorescence pattern starts to become bimodal, with two spatial lobes lying along the atomic beam trajectory ($z$). The separation of the scattering peaks increases with laser power, with a notable dip in fluorescence in the center (as seen in the image on the bottom of the inset in \ref{f.illustration}(c)). A sequence of true-color images obtained at 6 different laser powers is shown in Fig.~\ref{f.images}.

\section{Analysis}

\subsection{Model for Excited-State Population}

For fixed values of experimental parameters and for the specific angular momenta $J=1$ and $J'=0$ for our system, the steady-state $J'=0$ population, $\rho_{ee}$, can be calculated exactly, as shown in Ref.~\cite{berkeland}.  Although the local value of the laser intensity varies over the spatial extent of the mode, the steady-state population is reached fast enough that a fixed intensity can be used to calculate the excited-state population as a function of position; the $4s^2~{}^3{\rm P_0}$ state lifetime of 5~ns and atomic velocity of 500~m/s correspond to an atom traveling $\sim 10 \mu$m, a small fraction of the 2.5~mm laser mode, before the population reaches steady-state.

The solution for $\rho_{ee}$ in Ref.~\cite{berkeland} is obtained by solving the equation of motion for the atomic density operator with time evolution of the Zeeman sublevels $|J=1,m_j=\pm1\rangle$ due to the energy shift from the applied magnetic field.  The formalism in that work was developed to quantify the impact of an applied magnetic field (or laser-polarization modulation) on the dark-state that would otherwise result.  The magnetic field causes the lower-level sublevels to evolve in time at different rates, disrupting the steady-state solution in which no excited-state amplitude is generated.  Inserting our experimental parameters into their Eq.~(4.1) gives the following steady-state population in $J'=0$ as a function of Rabi frequency $\Omega$, magnetic-field frequency shift $\delta_B=\mu_B B /\hbar$ ($\mu_B$ is the Bohr magneton, and $B$ the magnitude of the magnetic field), and natural width of the excited state $\gamma/2\pi$:
\begin{equation}
\rho_{ee} = \frac{3}{4} \left(\frac{\Omega}{\gamma}\right)^2 \frac{1}{\frac{5}{2} - \left(\frac{\Omega}{\gamma}\right)^2\left(1 - 2  \left( \frac{\Omega}{4\delta_B}\right)^2- 2\left(\frac{4\delta_B}{\Omega}\right)^2\right)}. \label{e.Pee}
\end{equation}
This expression is for zero detuning and a linear polarization at an angle of $\pi/4$ with respect to the magnetic field direction, as in our system; inserting our specific magnetic field of $\delta_B=0.4\gamma$ yields
\begin{equation}
\rho_{ee} \approx \frac{3}{4} \left(\frac{\Omega}{\gamma}\right)^2\frac{1}{7.6 - \left(\frac{\Omega}{\gamma}\right)^2+ 0.8 \left(\frac{\Omega}{\gamma}\right)^4}. \label{e.Pee2}
\end{equation}

\begin{figure}
\includegraphics[width=0.5\textwidth]{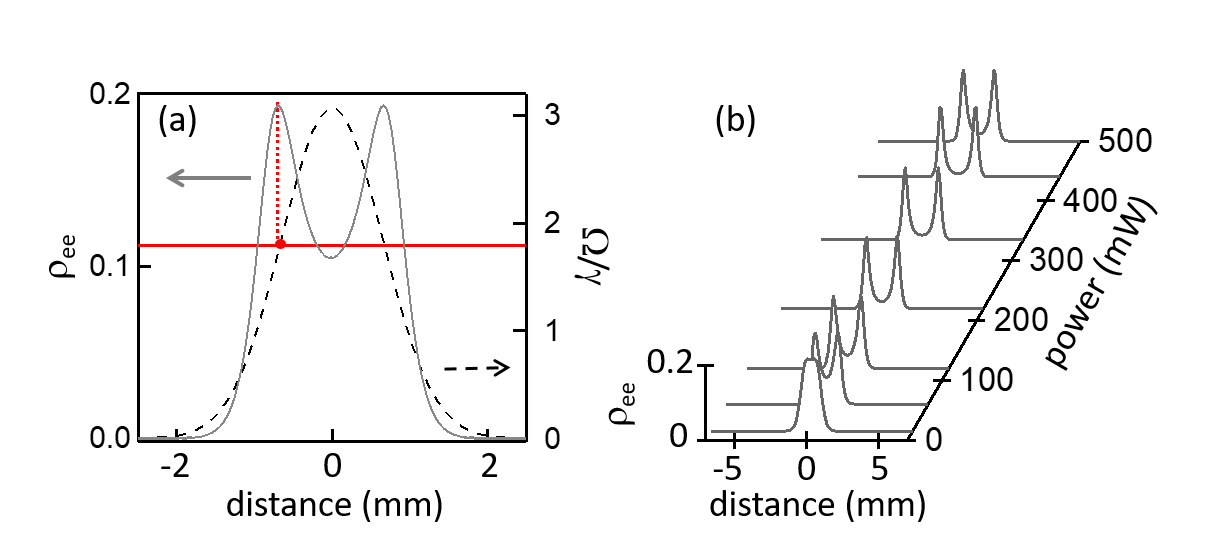}
\caption{(Color online.) (a) Calculated excited-state population (solid grey curve, left axis) for Gaussian intensity distribution (dashed black curve, right axis) and magnetic field value of $0.4 \gamma$. For low local intensities, the Rabi frequency associated with the optical interaction is not fast enough for the population to adiabatically follow the time evolution introduced by the magnetic field. The vertical (red) dashed line shows the point at which the Rabi frequency and Zeeman detuning are on par.  As the intensity increases further, population accumulates in the dark state at the expense of the excited state.  (b) Series of curves showing calculated excited-state population as a function of position ($z$) at the same laser powers as used for the data in Fig.~\ref{f.images}. These are 1D cross-sections of 2D solutions to Eq.~(\ref{e.Pee}).  The horizontal axis is scaled to $\pm 7~$mm to match the similar plot in Fig.~\ref{f.images}(b).} \label{f.model}
\end{figure}

The competition between laser and magnetic fields in determining the excited-state population is illustrated in Fig~\ref{f.model}.  In (a) we calculate the solution to Eq.~(\ref{e.Pee2}) for a Gaussian laser mode with waists of 1.3~mm along $z$ and 2~mm along $y$.  The total power chosen for this illustration is 40~mW.  The behavior of $\rho_{ee}$ corresponds to transitions between predominantly bright and predominantly dark atomic states.  The maximum excited-state population is about 0.25, half of the steady-state value for a two-state system driven to saturation. This maximum occurs when the laser intensity is given by $\Omega=1.8\gamma$; at higher intensities the time-evolution due to the optical field dominates the dephasing due to the magnetic field and a dark state begins to arise.  The value of the maximum excited-state population and the optical intensity at which the maximum population occurs depend on both the magnetic field strength and the total laser power.  In Fig~\ref{f.model}(b) a `waterfall' plot is shown for $\rho_{ee}$ for the same laser powers as used for the data in Fig.~\ref{f.images} and with the same horizontal scale as in that figure.  It can be seen that the peak fluorescence can occur outside the beam waist of the laser mode for large powers.

\subsection{Additional Details}

\begin{figure}
\includegraphics[width=0.5\textwidth]{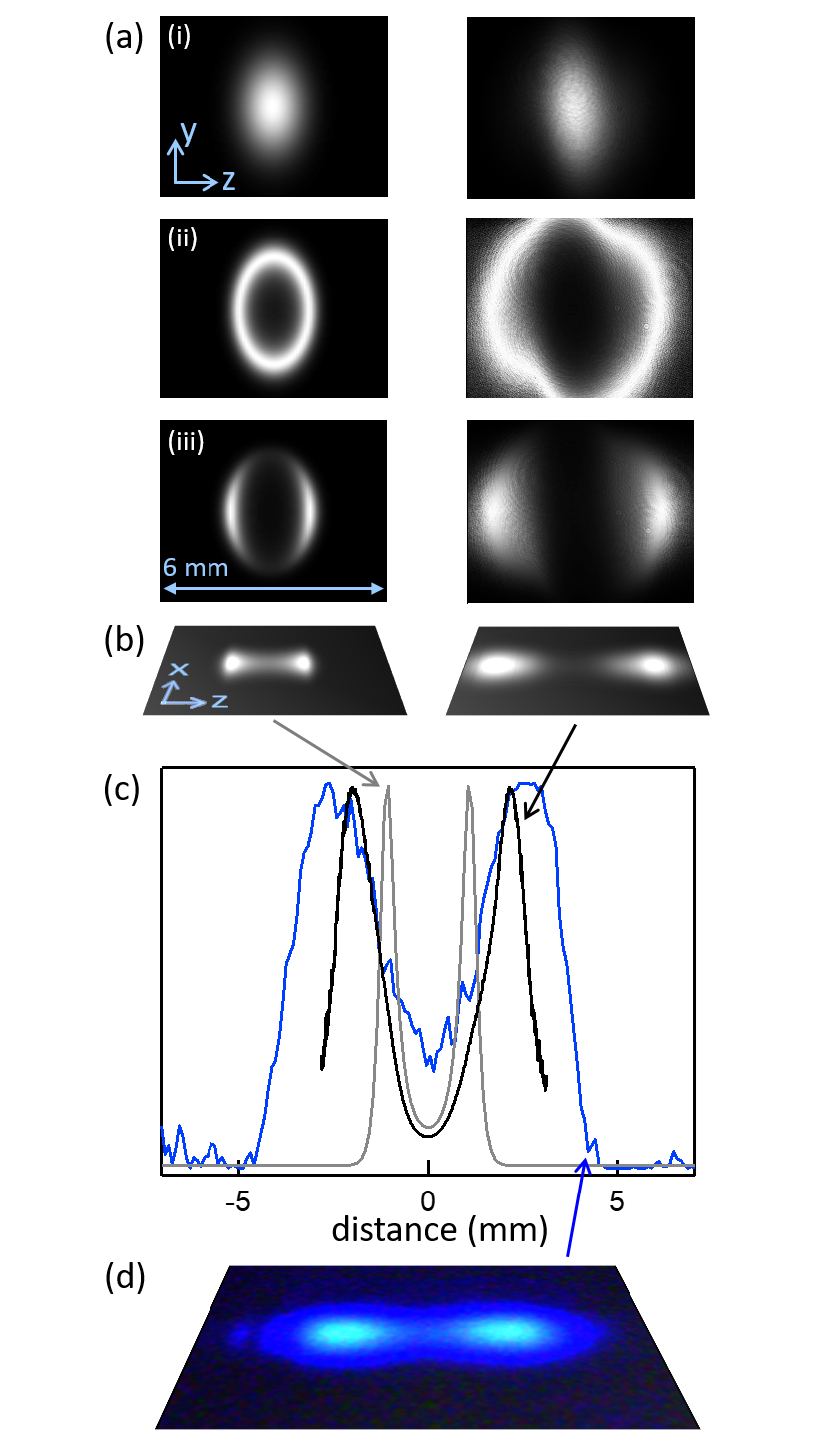}
\caption{(Color online.) Calculations of excited-state population (proportional to scattering rate) including additional experimental details.  (a) Series of images in $zy$ plane showing laser intensity distribution (i), corresponding excited-state population from Eq.~(\ref{e.Pee}) with 500~mW of laser power (ii), fraction of distribution that materializes due to limited extent of beam of atoms excited by 657~nm laser (iii), using calculated Gaussian distribution (left) and physical laser mode (right).  (b) Modeled fluorescence pattern observed with camera in $zx$ plane after integrating the image in (a)(iii) along $y$ and multiplying by Gaussian distribution of atoms along $x$. (c) Plot of 1D cross-section of observed fluorescence pattern (blue curve) and 1D cross-section of calculated distributions in (b) (dotted grey curve - Gaussian laser mode; solid black curve - physical laser mode).  The horizontal axis extends over 14~mm to account for observed features.  The calculations based on the physical laser mode extend only about 6~mm, the size of the detector on the profile meter. (d) Fluorescence image acquired with 500~mW of laser power.} \label{f.final}
\end{figure}

Calculating the excited-state population along a 1D slice of a Gaussian laser mode is a simplification compared to the actual experimental arrangement.  We develop the model further and apply it in Fig.~\ref{f.final} to the maximum power used in our system of 500~mW.  On the left in (a)(i) we show a modeled 2D laser intensity profile with waists along $z$ of 1.3~mm and $y$ of 2~mm, the values measured for our laser mode using a profile meter.  The (left) image in (a)(ii) is a plot of the solution for the excited-state population $\rho_{ee}$ for all points in this intensity distribution, and in (a)(iii) the fluorescence pattern from atoms traversing only a fraction of the laser mode is shown.  This is obtained by multiplying the image in (a)(ii) with a Gaussian distribution (waist of 1.5~mm) along $y$ representing the extent of the atoms in the beam excited by the 657~nm laser.  The size chosen for this waist is estimated from the size of the 657~nm mode at the location of the atoms plus the divergence of the atomic beam between the two interaction regions. The camera in our experiment observes fluorescence from below, collecting light from all atoms emitting. This aspect of the data acquisition is captured in (b), where the fluorescence integrated along $y$ is shown in the $xz$ plane.  The extent of the image along $x$ is modeled by using another Gaussian distribution to represent the extent of the atoms in the beam, this time along $x$; the spatial extent of the atomic beam in that direction is due to the spread of transverse velocities corresponding to Doppler shifts within $\pm250$~kHz of the 657~nm resonance.

A 1D cross-section of the left image in Fig.~\ref{f.final}(b) is plotted as the grey-dashed curve in (c).  The blue curve in the plot is the 1D cross section of a 500~mW fluorescence image, shown in the $xz$ plot in (d).  The location and width of the peaks of the calculated fluorescence distribution do not agree with observation as well as could be expected.

Better agreement is achieved between the data (blue curve) in the graph in (c) and the black curve.  The black curve corresponds to the 1D cross-section of the excited-state population distribution on the right in (b), which is calculated using the actual laser intensity distribution measured with a profile meter, shown on the right in (a)(i).  It can be seen that accounting for the deviation of the physical laser mode from a pure Gaussian results in far better agreement with observation.  The laser light used in the experiment is the output of a frequency doubled Ti:sapphire laser.  In order to investigate fluorescence patterns at the highest powers possible, we avoided spatial filtering or fiber coupling of the laser light, leaving us with a non-ideal spatial mode.

The black curve in Fig.~\ref{f.final}(c) is calculated from Eq.~(\ref{e.Pee2}) using measured experimental inputs and no free parameters, other than an overall scaling of signal size. The remaining discrepancy between the model and data, particularly regarding the width of the fluorescence peaks, is likely due to two sources of uncertainty. The first is the value of the magnification of the imaging system; calibration of the magnification could not be carried out at the position of the atoms inside the vacuum chamber, but rather only by using a scale at the nearest viewport.  An error in this scaling can change the width of the fluorescence peaks compared to the model.  The second source of uncertainty is the spatial variation of the bias magnetic field.  The coils used to create the field in the 431~nm interaction region are far from the ideal Helmholtz arrangement, resulting in a non-uniform field.  We have demonstrated that using a non-uniform magnetic field in the calculation can broaden the modeled fluorescence peaks.

\subsection{423~nm Transition}

The 431~nm transition provides a detection signal that is free from the background of ground-state atoms, {\em i.e.} those that do not get excited to the metastable $4s4p~{}^3{\rm P_1}$ state with the 657~nm laser.  Alkaline-earth frequency standards have more commonly used the $^1{\rm S_0}-{}^1{\rm P_1}$ transition for detection~\cite{australia, china, bruno}.  This cycling transition connects the ground state to a short-lived excited state, enabling photons to be scattered at a high rate.  Excitation of the ${}^1{\rm S_0}-4s4p~{}^3{\rm P_1}$ is measured as a reduction of the 423~nm fluorescence from the ground state atoms.  Because all ground state atoms (in the absence of 657~nm light) contribute to this cycling process, this transition is also useful for measuring atomic flux and general atomic-beam characterization.

\begin{figure}
\includegraphics[width=0.5\textwidth]{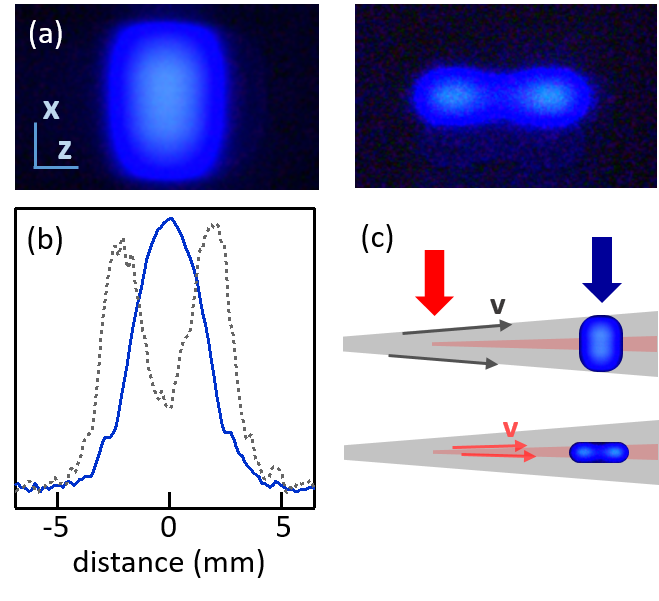}
\caption{(Color online.) (a) Fluorescence image from the $^1{\rm S_0}-{}^1{\rm P_1}$ transition, for 700~mW of 423~nm laser power, shown on the left. For comparison, a fluorescence image from the $4s4p~{}^3{\rm P_1}-4s^2~{}^3{\rm P_0}$ transition, for 500~mW of 431~nm laser power, is shown on the right. The scales of the two images are the same. (b)~Plot of 1D cross-section of 423~nm fluorescence pattern in (a) (solid blue curve), showing none of the suppression of scattering at the center as in the 431~nm transition; for the $J=0$ to $J'=1$, 423~nm transition, no dark state is expected to form. For comparison, the cross-section for the 431~nm $J=1$ to $J'=0$ transition is also shown (dashed grey line). (c)~The different extents of the fluorescence images along $x$ can be understood from the range of transverse-velocity Doppler shifts that are resonant with the 500-kHz (resolution-time limited) 657~nm interaction versus those that are resonant with the 34-MHz 423~nm interaction, as illustrated.  Ground-state atoms with any transverse velocity in the divergent atomic beam are resonant with the 423~nm laser, producing fluorescence for the entire width of the atomic beam (all $x$) (top). Only atoms originally excited by 657~nm light, with Doppler shifts from transverse velocities of $\pm 250~$kHz from resonance, subsequently scatter 431~nm; this smaller group of angles translates to a smaller spatial extent in $x$ of fluorescence (bottom).} \label{f.423}
\end{figure}

In calcium the $^1{\rm S_0} - {}^1{\rm P_1}$ transition has similar properties to the $4s4p~{}^3{\rm P_1}-4s^2~{}^3{\rm P_0}$ transition at 431~nm.  The transition occurs at a wavelength of 423~nm, has a natural width of $\gamma/2\pi=34$~MHz and a saturation intensity of 30~mW/cm$^2$.  We can look at the fluorescence patterns from this similar transition in the same detection region by replacing the 431~nm source with one at 423~nm.  The mode size and power for the lasers at the two wavelengths are similar.  Unlike the 431~nm transition, the fluorescence pattern from the 423~nm interaction shows a single fluorescence peak for all powers; no reduction in the center of the pattern is observed.  An example fluorescence image from this transition is shown in Fig.~\ref{f.423}.

There are two significant differences between excitation of the 423~nm and the 431~nm transitions that play a role in the $z$-dependence of the  fluorescence patterns.  The first is that the $J=1$ to $J'=0$ transition at 431~nm enables formation of a dark state via CPT, whereas the $J=0$ to $J=1'$ at 423~nm transition does not.  The other difference is that 431~nm excitation is part of a double resonance; only atoms first excited by 657~nm light can be subsequently driven by the 431~nm field. The aspect of double resonance that is important is the spatial selection of the subsequent 431~nm interaction imposed by the geometry of the first excitation.  Because the mode of the 657~nm laser is smaller (2~mm round) than that of the blue laser, the atoms excited by the 657~nm beam that are prepared to interact with the 431~nm light sample only a fraction of the spatial mode (see Fig.~\ref{f.illustration}(b)). This removes some of the averaging over Rabi frequencies that would occur if fluorescence from the entire spatial mode were collected.  (The fact the 423~nm fluorescence is collected from atoms sampling all of the spatial mode of the laser is likely the reason that the 1D cross-section in Fig.~\ref{f.423}(b) does not exhibit the expected saturation for that high intensity; the wings of the spatial mode contribute a fluorescence profile that does not exhibit saturation.)

For completeness, we point out that the difference in extent of fluorescence for 431~nm and 423~nm along $x$ stems from the different transverse velocity classes with Doppler shifts that are resonant with the excitation laser.  The broad 423~nm transition is resonant with the excitation laser for all transverse velocities in the diverging atomic beam. Since this laser couples to ground-state atoms, fluorescence is observed over the entire spatial extent of the atomic beam. The 657~nm transition is resonant with the excitation laser for a smaller range of transverse velocities, resulting in a small spatial extend along $x$ for the subsequent 431~nm fluorescence.  This is illustrated in Fig.~\ref{f.423}(c).

\section{Conclusions}

With enough experimental detail included, Eq.~(\ref{e.Pee}) reproduces the observed bi-modal fluorescence patterns with reasonable agreement.  This model then should enable optimization of the detection process in systems such as ours.  The amount of laser power available will determine the size of the magnetic field required to optimize the photon scattering rate, and the impact on coherence of the size and stability of the required magnetic field can be empirically investigated.  Additionally, elongation of the probe laser beam (along $z$) and corresponding design of the collection optics can be used to reduce peak intensity and avoid a dip in fluorescence while increasing the size of the region over which photons are scattered.  The best solution will be specific to the experimental design and constraints of a given system. Additionally, further refinement in applying the model could enable fluorescence images to be used to calibrate the strength of the applied magnetic field or laser intensity distribution {\em in situ}.  This could be useful, for instance, for measuring fairly large magnetic fields without the need to tune the laser frequency over tens of MHz in order to map out the Zeeman resonances.

In summary, we have observed competition between the laser intensity driving a CPT dark state and the magnetic field disrupting the CPT process along the intensity distribution of a laser mode.  
This competition can be present in state detection for atomic beam clocks, where it can impact the clock frequency if it leads to variations in the fluorescence signal---driven by variations in one of the competing fields---over timescales shorter than a measurement cycle~\cite{FM}.  These systematic frequency shifts can be predicted and mitigated by using analysis similar to that presented here.  Future investigation of the similar transition in strontium~\cite{bruno} or other atoms used in optical beam clocks would be of interest.

\section{Acknowledgements}

We are grateful for fruitful discussions with D.~Steck and J.~Hanssen.  Optical clock development at the U.S. Naval Observatory receives funding from the Office of Naval Research.


\begin{thebibliography}{10}

\bibitem{measurement}
Laser-induced fluorescence for measurement of the quantum state in atomic clocks is discussed in R. Wynands, S. Weyers, Metrologia {\bf 42}, S64 (2005) and A. D. Ludlow, M. M. Boyd, J. Ye, E. Peik, P. O. Schmidt, Rev. Mod. Phys. {\bf 87}, 637 (2015).

\bibitem{phillips}
W.~D.~Phillips, Rev. Mod. Phys. {\bf 70}, pp. 721-741 (1998).

\bibitem{arimondo}
E.~Arimondo, in Progress in Optics (Elsevier, Amsterdam, 1996) {\bf 35}, pp.257-354.

\bibitem{subrecoil}
A.~Aspect, E.~Arimondo, R.~Kaiser, N.~Vansteenkiste, and C.~Cohen-Tannoudji, Phys. Rev. Lett. {\bf 61}, pp. 826-829 (1988).

\bibitem{CPTclock}
M. Stahler, R. Wynands, S. Knappe, J. Kitching, L. Hollberg, A. Taichenachev, and V. Yudin, Opt. Lett. {\bf 27}, 1472-1474 (2002)

\bibitem{ion}
T.~Lindvall, M.~Merimaa, I.~Tittonen, and A.~A.~Madej, Phys. Rev. A {\bf 86}, 033403 (2012).

\bibitem{berkeland}
D.~J.~ Berkeland, M.~G.~Boshier, Phys. Rev. A {\bf 65}, 033413 (2002).

\bibitem{rms}
$\Omega$ refers to the ``rms'' Rabi frequency, {\em i.e.} the root mean square of the individual Rabi frequencies connecting the lower and upper sublevels.

\bibitem{ion2}
D. J. Berkeland, J. D. Miller, J. C. Bergquist, W. M. Itano, and D. J. Wineland, Phys. Rev. Lett. {\bf 80}, 2089 (1998).

\bibitem{ludlow}
J.~Olson, R.~W.~Fox, T.~M.~Fortier, T.~F.~Sheerin, R.~C.~Brown, H.~Leopardi, R.~E.~Stoner, C.~W.~Oates, and A.~D.~Ludlow, Phys. Rev. Lett. {\bf 123}, 073202 (2019).

\bibitem{Taylor}
J.~Taylor, B. Hemingway, J. Hanssen, T. B. Swanson, S. Peil, J. Opt. Soc. Amer. B {\bf 35}, 1557 (2018).

\bibitem{Lim}
J.~Lim, K.~Lee, J.~Ahn, Opt. Lett. {\bf 37}, 3378-3380 (2012).

\bibitem{Reetz}
M.~Reetz-Lamour, T.~Amthor, J.~Deiglmayr, M.~Weidemuller, Phys. Rev. Lett {\bf 100}, 253001 (2008).

\bibitem{width}
We observe the feature itself move in the $x$ direction when scanning the frequency of the 657~nm laser, indicating an atomic beam extent in $x$ considerably larger than the features we observe.

\bibitem{australia}
J. J. McFerran, A. N. Luiten, J. Opt. Soc. Am. B {\bf 27}, 277 (2010).

\bibitem{china}
H.~Shang, X.~Zhang, S.~Zhang, D.~Pan, H.~Chen, J.~Chen, Opt. Express {\bf 25}, 30459-30467 (2017).

\bibitem{bruno}
I. Manai, A. Molineri, C. Fréjaville, C. Duval, P. Bataill, R. Journet, F. Wiotte, B. Laburthe-Tolra, E. Maréchal, M. Cheneau, M. Robert-de-Saint-Vincent, arXiv, 1910.11718 (2019).

\bibitem{FM}
Typically measurement on an atomic beam is carried out by frequency modulating the clock laser (657~nm laser in this work) and demodulating the detection signal, and the modulation period sets the timescale for the measurement cycle.


\end{thebibliography}
\end{document}